\documentclass[aps,prl,twocolumn,showpacs,superscriptaddress]{revtex4-1}
\usepackage{amsmath}
\usepackage{epsfig}
\usepackage{color}
\usepackage{endnotes}
\usepackage{natbib}
\usepackage[colorlinks,breaklinks]{hyperref}
\usepackage{nicefrac}
\usepackage{bm}% bold math
\usepackage{dcolumn}
\usepackage{graphics}% Include figure files
\usepackage{epstopdf}
\usepackage{amsmath}
\usepackage{amssymb}
\usepackage{color}
\usepackage{changes}
\newcommand{\ket}{\rangle}

\newcommand{\bs}[1]{\ensuremath{\boldsymbol{#1}}}

\newcommand{\be}{\begin{equation}}
\newcommand{\ee}{\end{equation}}
\newcommand{\bea}{\begin{align}}
\newcommand{\eea}{\end{align}}
\newcommand{\beqa}{\begin{eqnarray}}
\newcommand{\eeqa}{\end{eqnarray}}

\newcount\colveccount
\newcommand*\colvec[1]{
        \global\colveccount#1
        \begin{pmatrix}
        \colvecnext
}
\def\colvecnext#1{
        #1
        \global\advance\colveccount-1
        \ifnum\colveccount>0
                \\
                \expandafter\colvecnext
        \else
                \end{pmatrix}
        \fi
}
\newcommand{\rvec}{\bs{r}}
\newcommand{\kvec}{\bs{k}}
\begin{document}

\title{Contact formalism for coupled channels}

\author{Ronen Weiss}
\affiliation{The Racah Institute of Physics, The Hebrew University, 
             Jerusalem, Israel}
\author{Nir Barnea}
\email{nir@phys.huji.ac.il}
\affiliation{The Racah Institute of Physics, The Hebrew University, 
             Jerusalem, Israel}

\date{\today}

\begin{abstract} 
The contact formalism, a useful tool for analyzing short-range correlations,
is generalized here for systems with coupled channels, such as in nuclear physics.
The relevant asymptotic form is presented and contact matrices are defined.
Generally, for the case of two coupled channels, two two-body functions are included
in the asymptotic form, resulting a $2 \times 2$ contact matrix.
Nevertheless, it is shown that if the coupling terms of the potential 
are very weak or very strong,
only a single two-body function is needed, resulting a single contact. This universal result
is directly relevant to nuclear systems, and provides a
theoretical explanation for the fact that proton-neutron short-range
correlations can be described using the single
bound-state deuteron wave function.
It is achieved
by applying an appropriate boundary condition on the two-body functions. 
This boundary condition can be interpreted as a mean field
potential imposed on the correlated pair due to the residual system.

\end{abstract}

% 67.85.-d - ultracold gases
% 05.30.Fk - Fermion systems and electron gas
% 25.20.-x - Photonuclear reactions

\maketitle
%=============================================================================
% Introduction
%=============================================================================
{\it Introduction --}
The contact formalism is a relatively new technique
for analyzing short-range correlations (SRCs) in quantum systems.
Originally, the contact was devised by Shina Tan 
to describe systems consisting of two-component fermions,
fulfilling the zero-range condition \cite{Tan08}.
%, in which
%the interaction range is much smaller than any other
%length scale of the system, such as the scattering length or the 
%average interparticle distance.
Under this assumption, different properties of the system 
were related to a single variable, the contact \cite{ Bra12}.
Many of these relations were also
verified experimentally in ultra cold atomic systems
\cite{SteGaeDra10,SagDraPau12,ParStrKam05,WerTarCas09,KuhHuLiu10}.
The success of this theory led to concentrated attempts
to generalize it to other physical systems. The first studies in this
direction generalized the formalism to bosonic, and
mixed fermions systems
\cite{ComAlzLey09,BraKanPla11,WerCas12Bosons,
WilJin_BEC_Exp,MakJin_exp_bos,SmiBraPla_Exp_analysis_Bos},
and also to different dimensions and non-trivial geometries
 \cite{ComAlzLey09,WerCas12Bosons,ValZinKLa11,
ValZinKLa12,WerCas12fermions,BarZwe11,Hofmann12,
LanBarBra12,Hofmann11}.
Recently, a $p$-wave contact was defined and utilized to describe
systems with a resonant zero range $p$-wave interaction
\cite{YosUed_pwave,YuThyZha15_pwave,p_wave_exp}.
%The relevant contact relations were generalized,
%and few of them were also verified experimentally \cite{p_wave_exp}.
%Recently, considering various partial waves 
%the idea of contact spectrum was also introduced \cite{WeiBazBar15a}, 
%\red{and utilized} \cite{HeZhou16_spectrum}.

The contact formalism was also generalized and utilized to study nuclear systems
\cite{WeiBazBar15,WeiBazBar15a,WeiBazBar16,WeiPazBar,Hen15,BaoJun16,WeiHen17}.
Nuclear systems do not obey the zero range condition,
and as a result few significant changes had to be made. Mainly,
the contribution of all partial waves should be considered,
not only the $s$-wave contribution, and 
the known zero-range two-body (2B) functions, which are
an integral part of the contact theory, should be replaced with unknown
or model dependent functions. Accordingly, the nuclear contact matrices were defined.
Using these matrices, new relations between different nuclear quantities, such as
the photoabsorption cross section and momentum distributions,
were derived and verified
\cite{WeiBazBar15,WeiBazBar15a,WeiBazBar16,ciofi_new,WeiPazBar,WeiHen17}.
Similar contact matrices and contributions from different partial waves
were also considered recently to describe SRCs in other systems
\cite{HeZhou16_spectrum,ZhaHeZhou16,YosUed16}.

In this work we wish to take the contact formalism one step
further, adapting it to describe systems dominated by coupled channels,
situation ignored in previous studies.
In order to make our goal clearer, we consider a system
of two-component fermions with spin $s=1/2$, interacting via
non-central 2B force that couples different orbital angular momentum $\ell$ and
spin $s$ channels. This force does not necessarily need to
fulfill the zero-range condition. Looking on the 2B system, it can be characterized
using the quantum numbers $(\ell,s)j\,m$, 
where $j$ and $m$ are the total angular momentum and its projection. 
%The meaning of
%coupled channels here is that the 2B potential
%mixes between the different quantum numbers
%$\ell$ and $s$ , for example. 
As the potential mixes different $\ell,s$ channels, 
an eigenstate of the 2B system does not necessarily have a well defined 
$\ell,s$ values.
The nuclear potential is an example for such an interaction, as, due to pion exchange,
the deuteron is a bound 2B state with $j=1$ and $s=1$ but with both $s$-wave
and $d$-wave components. 
%In the recent generalizations of the contact formalism
%to nuclear systems, this point was not taken into consideration. 

%In the following we first present the 
%changes necessary for applying the contact formalism
%to systems with coupled channels. Then we discuss the difference
%between strong and weak coupling and its implication to nuclear systems,
%and show that in the weak and strong coupling limits, only a single contact
%dominates.

%=============================================================================
% coupled channels contact formalism
%=============================================================================
{\it Coupled channels contact formalism --}
Consider a system of bosons,
interacting via zero range $s$-wave potential.
In such system, when two particles
approach each other,
the total wave function $\Psi(\rvec_1,\ldots,\rvec_N)$ takes the form
\be
  \Psi  \xrightarrow[r_{ij}\rightarrow 0]{}\varphi(\bs{r}_{ij})
           A(\bs{R}_{ij},\{\bs r_k\}_{k\neq i,j})\;.
\ee
where, $N$ is the number of particles,
%$\bs{r}_k$ are the single particle coordinates, and
$\bs{r}_{ij}=\bs{r}_i-\bs{r_j}$ and
$\bs{R}_{ij}=(\bs{r}_i+\bs{r_j})/2$,
% are the pair's relative
%and center of mass coordinates, respectively, 
$\varphi(\bs{r}_{ij})$ is the universal zero-energy solution of the
%time-independent 
Schr\"{o}dinger
equation of the 2B system, and $A$ is a {\it regular} function describing  
the dynamics of all other degrees of freedom.
For an $s$-wave interaction and under the zero-range assumptions,
 $\varphi=\left(1/r_{ij}-1/a\right)$, where $a$ is
the scattering length.
This factorization is closely related to the 
operator product expansion \cite{OPE_wilson}, which was used to derive
some of the known contact relations \cite{Braaten_OPE, Hofmann11,BarZwe11}.

When considering also higher partial waves, if the different 
channels are not coupled, the asymptotic form
becomes \cite{WeiBazBar15a, HeZhou16_spectrum}
\be \label{non_coupled_asymp}
  \Psi  \xrightarrow[r_{ij}\rightarrow 0]{}
	\sum_{\alpha} \varphi_\alpha(\bs{r}_{ij})
           A_\alpha(\bs{R}_{ij},\{\bs r_k\}_{k\neq i,j})\;.
\ee
Here, the sum over $\alpha$ indicates the different channels.
%For example, in a system of spinless bosons or one-component
%fermions \cite{p_wave_exp},
%we will have $\alpha=\ell\,m_\ell$ \cite{ZhaHeZhou16},
%which are the quantum numbers of
%the orbital angular momentum.
In a system of two-component
fermions $\alpha=(\ell,s)j\,m$ as discussed 
above. 
%It is important to recall that $\varphi_\alpha$
%should be the zero-energy solution of the 2B
%Schr\"{o}dinger equation with the relevant quantum numbers.
%Consequently, this asymptotic form makes sense only
%if the quantum numbers indicated by $\alpha$ are good quantum number.
%%Otherwise, $\varphi^\alpha(\bs{r}_{ij})$ are not well defined.
%As mentioned above, for coupled channels the quantum numbers
%indicated by $\alpha$  are not necessarily
%good quantum numbers. Thus, this asymptotic form is not suitable
%for the case of coupled channels and the first step for generalizing the contact
%formalism to such systems is to understand the appropriate asymptotic
%form. 
When considering coupled channels, the quantum numbers
indicated by $\alpha$  are not necessarily
good quantum numbers, and we should seek a different form for the universal
part of the asymptotic wave-function.

To this end, we can continue with the example
of two-component fermions with the quantum numbers
$(\ell,s)j\, m$, and  
consider an asymptotic 2B state composed of
two coupled channels, given by
$|\alpha\ket=|(\ell_\alpha,s_\alpha)j\, m\ket$ and
$|\beta\ket=|(\ell_\beta,s_\beta)j\,m\ket$ with
$\ell_\alpha \neq \ell_\beta$ (generalization to $n$
channels is straightforward). 
In order to find the suitable
asymptotic form, we need to understand what are
the zero-energy solutions. In this
case the Schr\"{o}dinger equation becomes a set of two coupled equations,
which results two independent solutions. 
%Any solution of these
%coupled equations can be written as a linear combination of these two solutions.
%If these two channels were not coupled, then we would also have two
%independent solutions, one for each channel. Since the channels are coupled,
%each of the two solutions includes component of both channels.
These solutions have the following form
\be \label{coupled_sol}
\varphi^a(\bs{r}) = \varphi^a_\alpha(r) |\alpha \rangle +
 		 \varphi^a_\beta(r) |\beta \rangle , 
\ee
%where $|\alpha \rangle = [Y_{\ell_\alpha}(\hat{r}) \otimes \chi_{s_\alpha}]^{j_\alpha,m_\alpha}$,
%$|\beta \rangle = [Y_{\ell_\beta}(\hat{r}) \otimes \chi_{s_\beta}]^{j_\beta,m_\beta}$,
%$Y_{\ell m_\ell}$ are the spherical harmonics, $\chi_{s m_s}$ are the spin functions and
%$\bs{r}$ is the relative coordinate. 
where $a=1,2$ indicates the two independent solutions
that differ only by the functions $\varphi^a_\alpha$ and $\varphi^a_\beta$.
We notice that each of these solutions is a mixture of both channels,
$|\alpha \ket$ and $|\beta \ket$.

Now it becomes clear how the asymptotic wave function should look like.
If we consider only these two coupled channels, it will take the form
\be \label{coupled_asymp}
\Psi \xrightarrow[r_{ij}\rightarrow 0]{}
	\sum_{a=1,2} \varphi^a(\bs{r}_{ij})
           A^a(\bs{R}_{ij},\{\bs r_k\}_{k\neq i,j})\;.
\ee
Notice that it is a different asymptotic form then the non-coupled
case, given by Eq. \eqref{non_coupled_asymp}, since Eq.
\eqref{coupled_asymp} includes two different functions
$\varphi^1_\alpha(r)$ and $\varphi^2_\alpha(r)$,
and each of them is generally coupled to a different $A^a$
function.
For simplicity, we ignore the summation over $m$ required
if $\Psi$ has a well-defined total
angular momentum $J$. It is simple to make the necessary changes
(see Ref. \cite{WeiBazBar15a}),
and it does not affect the conclusions of this paper.

Based on the above asymptotic form, we can define
the contacts for the case of coupled channels. A matrix
of contacts should be defined
\be \label{coupled_contacts}
C^{ab}=16\pi^2 \frac{N(N-1)}{2} \langle A^a | A^b \rangle
\ee
where $a,b=1,2$. We have
defined here a $2 \times 2$ matrix of contacts, for the case
of two coupled channels. It is important to include the
non-diagonal element of the matrix, since $A^1$ and $A^2$ are not
necessarily orthogonal. 
A matrix of contacts was already defined in previous studies
\cite{WeiBazBar15a,ZhaHeZhou16,YosUed16},
in which it was implicitly assumed
that asymptotically the potential does not couple different channels.
Nevertheless, the off-diagonal terms of the matrix can still be important.
Here we consider the case of a potential that
couples between channels.

%%% We note that the above definition can be easily generalized to
%%% the case of $n$ coupled channels, which will lead to an $n \times  n$
%%% contact matrix. It can also be expanded to include different
%%% sets of coupled channels together, instead of the one set we have considered here.
%%% It will simply result more contacts, where one should also define non-diagonal
%%% contact elements that come from the different sets.

%%% Coming back to our simplified case, described by Eqs. 
%%% \eqref{coupled_asymp} and \eqref{coupled_contacts},
%%% we have said that mathematically it seems that we need two different functions
%%% to describe the asymptotic form in Eq. \eqref{coupled_asymp}, indicated by
%%% $i=1,2$. An interesting question is if indeed the two functions are needed
%%% or that in some scenarios one function is sufficient.
%%% In order to try and answer this question we will first 
%%% adapt few of the known contact relations
%%% to this case of coupled channels. It will allow us to see what are the differences
%%% between one or two functions in the asymptotic form.

Before analyzing the universal wave-function \eqref{coupled_sol} let us 
recast some contact relations using the asymptotic form \eqref{coupled_asymp}.
We will focus here on the momentum and density distributions.
The single-particle momentum distribution, $n(k)=\int d\hat{k} n(\kvec)$,
describing the probability to find a particle with momentum $k$, is given 
asymptotically by
\be \label{1b_mom}
n(k) \xrightarrow[k \rightarrow \infty]{}
\sum_{a,b=1,2} \left( \tilde{\varphi}^{a*}_\alpha(k)\tilde{\varphi}^{b}_\alpha(k)
+  \tilde{\varphi}^{a*}_\beta(k)\tilde{\varphi}^{b}_\beta(k)
 \right) \frac{2C^{ab}}{16\pi^2},
\ee
where 
%$n(\kvec)$ is normalized to the number of particles in the system $N$, and
$\tilde{\varphi}^a_\alpha(k)$
is the Fourier transform of $\varphi^a_\alpha(r)$, and 
%we have used the normalization
$\int \frac{d\kvec}{(2\pi)^3} n(\kvec)=N$.
%We have used here the orthogonality of the spherical harmonics functions
%under the integration over $\hat{k}$.
Similarly, the two-particle momentum distribution
$F(k)=\int d\hat{k} F(\kvec)$, which describes the probability to find a 
particle pair with relative momentum $k$, 
%integrated over the center of mass momentum,
is given by
\be \label{2b_mom}
  F(k) \xrightarrow[k \rightarrow \infty]{}
    \sum_{a,b=1,2} \left( \tilde{\varphi}^{a*}_\alpha(k)\tilde{\varphi}^{b}_\alpha(k)
   + \tilde{\varphi}^{a*}_\beta(k)\tilde{\varphi}^{b}_\beta(k)
                 \right) \frac{C^{ab}}{16\pi^2}.
\ee
%$F(k)$ is normalized to the number of pairs in the system.
The derivation of these two relations is very similar to the
derivation presented in \cite{WeiBazBar15a}.
It is also very simple to derive the asymptotic
probability to find a pair of particles with relative distance 
$r$, which is denoted by $\rho(r)=\int d\hat{r} \rho(\rvec)$.
In this case the relation is
\be \label{density}
\rho(r) \xrightarrow[r \rightarrow 0]{}
\sum_{a,b=1,2} \left( \varphi^{a*}_\alpha(r) \varphi^{b}_\alpha(r)
+  \varphi^{a*}_\beta(r) \varphi^{b}_\beta(r)
 \right) \frac{C^{ab}}{16\pi^2}.
\ee
$\rho(r)$ and $F(k)$ are normalized to the number of pairs.

Now we can see that if we have two different functions
$\varphi^a$, $a=1,2$, in the asymptotic form (Eq. \eqref{coupled_asymp}),
then the last three relations
% Eqs. \eqref{1b_mom}, \eqref{2b_mom} and \eqref{density},
are quite complicated. For example, if we will
compare two different eigenstates (with the same underlining
interaction), $\Psi_1$ and $\Psi_2$,
and look on the ratio of the two momentum distributions,
$n_1(k)/n_2(k)$, where $n_i(k)$ corresponds
to $\Psi_i$, there is no reason that this ratio will obtain a constant value
for high momentum. This is because the values of the different
four contacts $C^{ab}$ can be different for each state, and the
$k$-dependence will not generally disappear. The same argument
is relevant also for $F(k)$ and $\rho(r)$.
On the other hand, if there is only one asymptotic function,
then these ratios must have a constant
value for high momentum or small distances.
%This is because we will be left
%with only one term in Eqs. \eqref{1b_mom}, \eqref{2b_mom}
%and \eqref{density}.

The remaining question is whether two functions are indeed required in the
asymptotic form. We will first focus on nuclear systems,
and then use our new insights to try and provide a general answer.

%=============================================================================
% Nuclear system as an example
%=============================================================================
{\it Nuclear systems -- }
SRCs are known to play an important role
in nuclear physics. For example, a high-momentum
tail originated by SRCs was identified for $k>k_F\approx 1.26$ fm$^{-1}$
\cite{WirSchPie14,HenSci14,AlvCioMor08,ArrHigRos12,Fomin12}.
See also \cite{Cio15, Hen_review}.
Nuclear systems are more complicated then the simple case discussed above,
since they include both protons and neutrons, each with a spin degree of freedom.
Here, we will focus on proton-neutron (pn) pairs. The main
channels contributing to SRCs for pn pairs are the two coupled channels
$\alpha=(\ell_\alpha=0,s_\alpha=1,j=1,m)$ and
$\beta=(\ell_\beta=2,s_\beta=1,j=1,m)$
with isospin $T=0$.
%The single-particle momentum distributions are affected also
%by proton-proton or neutron-neutron pairs. Thus, we will focus
%on the $pn$ momentum distribution $F_{pn}(k)$ and density
%$\rho_{pn}(r)$, which are affected only by $pn$ pairs.

In the study of $pn$ SRCs in nuclear systems there are many 
indications that they can be described using the single $T=0$ deuteron
function to a good approximation. The deuteron is the 2B bound state in the above
mentioned $j=1$ channels. For example, in the analysis
of electron-scattering experiments, it is shown that for
kinematics sensitive to SRCs, the cross section becomes approximately
proportional to the deuteron cross-section \cite{Fomin12,ArrHigRos12,FraSar93}.
Similar picture is obtained from analysis of numerical calculations.
Using available numerical data
\cite{WirSchPie14}, it was shown in Ref. \cite{WeiBazBar15a}
that the $pn$ momentum distribution of the available nuclei,
is approximately a multiplication of the deuteron momentum distribution,
for $k>4$ fm$^{-1}$. It was also observed for heavier
nuclei using different numerical methods \cite{ciofi_new}.
Recently, the one-body momentum distribution $n(k)$
was reproduced using the contact formalism
for $k$ larger than the Fermi momentum \cite{WeiHen17}.
The main contribution to $n(k)$ comes from the deuteron channel, 
using the single bound state wave function.

All of these indicate that for some
reason only one function is needed in the asymptotic form
when a proton gets close to a neutron, with in the $T=0$ deuteron
channel.
We notice that there is a wide agreement in the literature
on the fact that the deuteron $T=0$ channel is the dominant channel 
in the description of nuclear SRCs, due to the nuclear tensor force
\cite{Piasetzky06,Subedi08,Baghdasaryan10,
Korover14,HenSci14,Schiavilla07,AlvCioMor08}.
Nevertheless, it was not realized that generally two independent 
functions in the deuteron channel are required and 
thus the deuteron channel dominance does not explain why nuclear
SRCs can be approximately described using only the bound-state deuteron wave
function.
%it is not understood why in practice a single asymptotic
%function is sufficient.

For a bound state in the case of two coupled channels,
a single solution is obtained due to the additional boundary condition at
$r \rightarrow \infty$.
Perhaps some boundary condition should also be applied
to the 2B zero-energy asymptotic wave functions,
describing the motion of the pair inside a nucleus.

%The first step in calculating
%the wave function of the deuteron
%for a given nuclear potential, is to calculate two independent
%solutions for the time-independent Schr\"{o}dinger equation. The second step
%is to find the correct linear combination that obeys the boundary condition
%at $r \rightarrow \infty$ of a bound state.
%Due to this additional boundary
%condition, a specific linear combination of the two independent solutions
%is chosen, and we are left with a single solution.

The functions $\left\{ \varphi^a(\bs r) \right\}_{a=1,2}$ are 
given in Eq. \eqref{coupled_sol}, and can also be written
in a vector form, such that a general solution will be written as a linear combination
\be
\varphi(\bs r) = c_1 \colvec{2}{\varphi^1_\alpha(r)}{\varphi^1_\beta(r)}+
c_2 \colvec{2}{\varphi^2_\alpha(r)}{\varphi^2_\beta(r)}.
\ee
The solutions are defined by the boundary conditions at $r=0$,
$\frac{d^{(\ell_\alpha+1)}}{dr^{(\ell_\alpha+1)}}(r \varphi^1_\alpha)=1$,
 $\frac{d^{(\ell_\beta+1)}}{dr^{(\ell_\beta+1)}}(r \varphi^1_\beta)=0$,
and
$\frac{d^{(\ell_\alpha+1)}}{dr^{(\ell_\alpha+1)}}(r \varphi^2_\alpha)=0$,
$\frac{d^{(\ell_\beta+1)}}{dr^{(\ell_\beta+1)}}(r \varphi^2_\beta)=1$.
$c_1$ and $c_2$ are free coefficients.
The simplest boundary condition is
of a spherical box with radius
$R$. It implies that the wave function must be zero at $r=R$,
%which means $\varphi(R)={\bf 0}$,
as a vector. 
Without such a boundary condition we have two independent solutions
for each positive energy. With this boundary condition, we will get
quantization of the positive energies, and for each allowed energy
there is only one solution characterized by the value of the ratio
$\eta \equiv c_2/c_1$.

\begin{figure}\begin{center}
\includegraphics[width=8.6 cm]{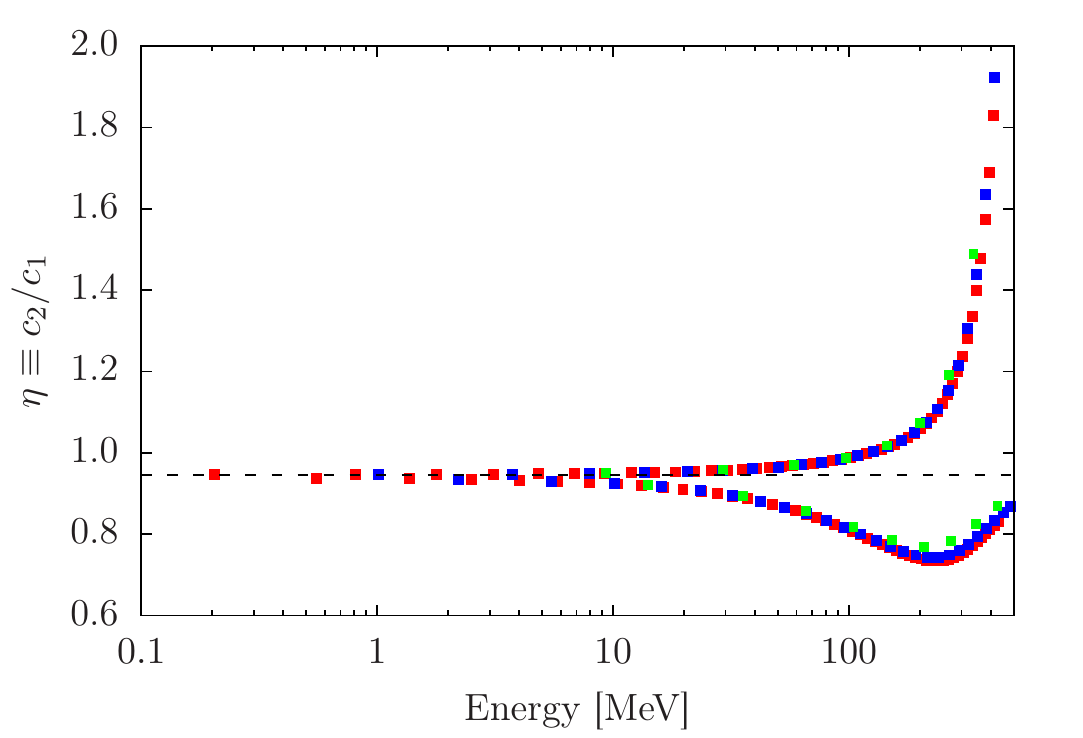}
\caption{\label{ratio_VS_energy} 
The ratio $\eta \equiv c_2/c_1$ vs. the allowed energies
for different box radii $R$. The red,
blue and green dots are
for $R=50, 25, 10$ fm, correspondingly.
%We can see that for
%small enough energies the ratio goes to a constant value
%independently of R.
}
\end{center}\end{figure}

In order to see the implications of such a boundary condition
we will use the AV18 two-nucleon potential \cite{av18}, and solve numerically
the 2B Schr\"{o}dinger equation.
As expected, without the external boundary condition, for any positive energy
we get two independent solutions $\varphi^1(\bs r)$ and $\varphi^2(\bs r)$. 
We note that for small enough energies and small enough distances,
these solutions do not depend on the energy.
When including the boundary condition, we can find the allowed
energies and the corresponding values of the ratio $\eta$
for each energy.
In Fig. \ref{ratio_VS_energy}, we present the values of this ratio for different energies
and using different values for the radius of the box $R$.
We stress once again that the value of $\eta$
determines the resulting 2B function  $\varphi(\bs r)$ for small
enough energies and small enough values of $r$. We can see in Fig.
\ref{ratio_VS_energy} that for
energies smaller than about $30$ MeV, the value of this ratio is approximately constant
independent of the value of $R$. This is quite a surprising
result, and it means that the spherical box somehow "chooses" the same linear
combination independently of the radius of the box or the energy of the
2B state. 

\begin{figure}\begin{center}
\includegraphics[width=8.6 cm]{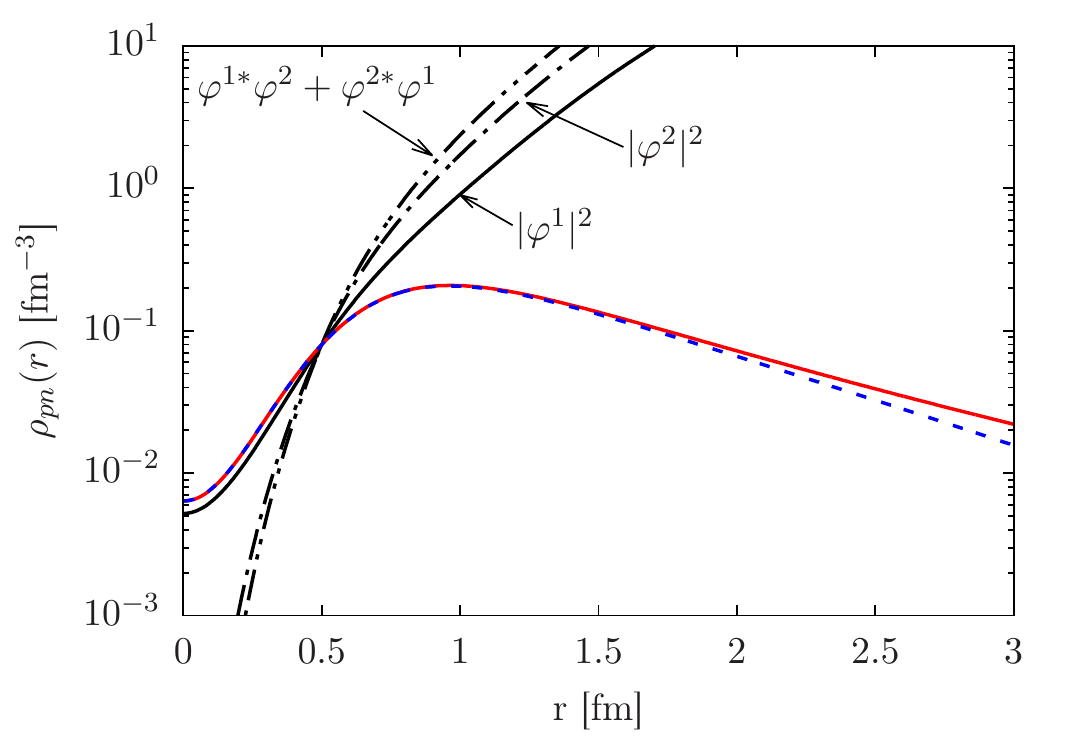}
\caption{\label{rho_pn} 
The deuteron density $\rho_{pn}(r)$ 
calculated using the exact bound state solution
(solid red line), is compared to $\rho_{pn}(r)$
calculated using positive energy solutions, Eq.
\eqref{density}. The dashed blue line 
corresponds
to the lowest allowed energy in a box
of radius $R=50$ fm ($E=0.20$ MeV).
The black lines
correspond to the $a=b=1$ term (continuous),
the $a=b=2$ term (dashed-dotted) and the mixed term
(dashed-double dotted) of Eq. \eqref{density}. All the curves are normalized
to a common value at $r=0.5$ fm.
}
\end{center}\end{figure}

These results can explain
why in nuclear systems we can describe the $T=0$ 
deuteron-channel SRCs using a single
function. Since we don't have a reason to prefer a
specific radius for the box, it is important that the chosen function does not depend 
on the radius. In order to make sure that this is indeed the correct
boundary condition, we can check if the chosen function $\varphi(\bs r)$ 
can describe the deuteron wave function. 
In Fig. \ref{rho_pn} we present $\rho_{pn}(r)$ as calculated directly
using the deuteron wave function for AV18 and using
a single positive energy solution $\varphi(r)$ in the presence of a 
spherical box. We can see that 
indeed these two quantities coincide for small distances.
Thus we conclude that the box boundary condition not only lead
to a single asymptotic function, but this function is also the deuteron's asymptotic
function and is thus suitable to describe $pn$ SRCs for heavier nuclei.
We can also see in Fig. \ref{rho_pn} that using arbitrary
positive-energy solutions,
without the box boundary condition,
the deuteron density $\rho_{pn}(r)$ cannot be reproduced.
Only the specific combination
determined by the box is suitable.

We note that the hard-wall boundary condition of the box 
can be replaced by a softer boundary condition.
For example, if we include an external wide harmonic-oscillator potential,
instead of the the box boundary condition,
we still obtain similar results.
Thus, this new boundary condition
can be interpreted as the effect of the residual particles
on the SRC pair, through a mean field potential imposed
on this pair. The exact details of this potential are not important.

This gives an explanation for the fact that
$np$ SRCs can be described
using only the deuteron wave function as leading order
approximation, without a second function for these
coupled channels. Theoretical explanation to 
this phenomena has not been presented in previous
studies.
Thanks to the nuclear example we have been able to identify the
necessary boundary condition for the 2B functions appearing in the
asymptotic form.  We can now in principle apply it to different systems
and study the implications of this new boundary condition.

\begin{figure}\begin{center}
\includegraphics[width=8.6 cm]{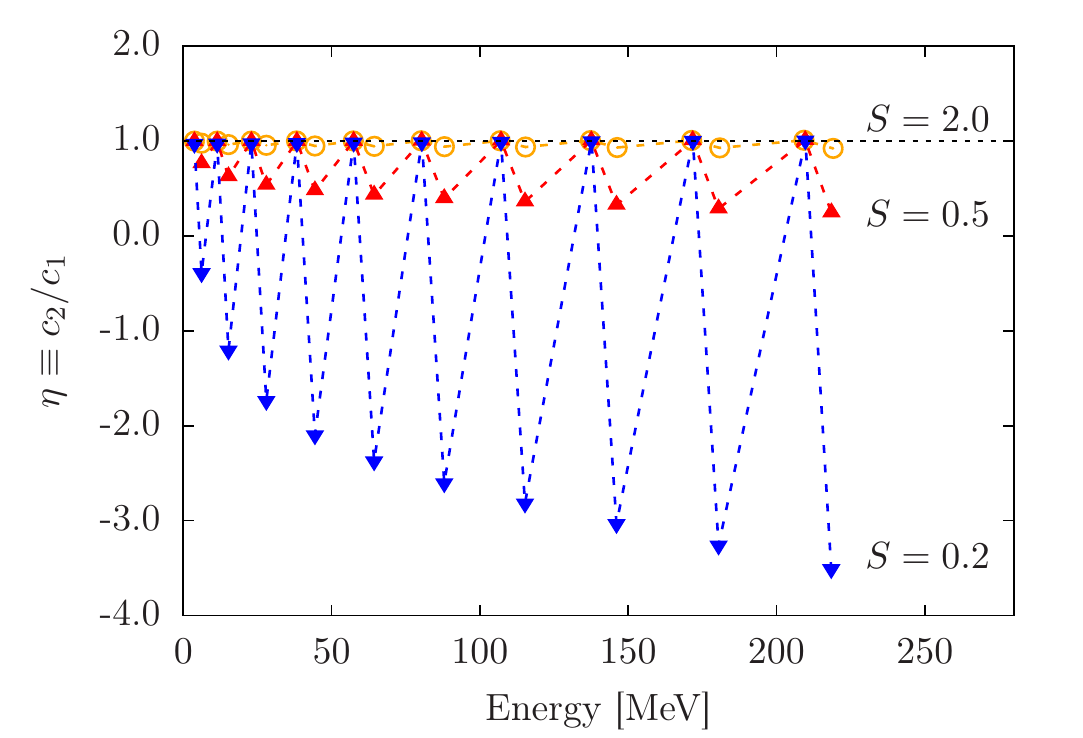}
\caption{\label{strngth} 
The ratio  $\eta(E) \equiv c_2(E)/c_1(E)$  in a box with radius $R=15$ fm
for different values of $S$, using
the simple potential given by Eqs. \eqref{Gauss1} and \eqref{Gauss2}. We used
$\Lambda^2=3$ fm$^{-2}$ and a fixed scattering length $a=10$ fm.
The blue down-pointing triangles,
red up-pointing triangles, and orange circles
are for $S=0.2, 0.5, 2$, respectively.
}
\end{center}\end{figure}

%=============================================================================
% General potential
%=============================================================================
{\it General potential -- }
To this end,
we consider a simpler potential. We will use the
$pn$ nuclear example in the deuteron channels as before,
but with a potential that has the following components
\be \label{Gauss1}
V_{\alpha\alpha}=V_{\beta\beta}=-V_0 \exp(-\Lambda^2 r^2)
\ee
\be \label{Gauss2}
V_{\alpha\beta}=V_{\beta\alpha}=- S V_0 \exp(-\Lambda^2 r^2)
\ee
where S represents the strength of the coupling terms. 
If $S=0$, the two channels are actually not coupled.
We can numerically solve the Schr\"{o}dinger equation for this potential
in a box. We want to see if again
we get a single solution or if in some cases we need two different functions.
We first choose a radius $R$ for the spherical box.
Then, for a given strength $S$, we tune $V_0$ to produce
a constant scattering length, and calculate the value
of $\eta$ as a function of the allowed energies.
The results are presented in Fig. \ref{strngth}.
We can see that as the coupling of the channels ($S$)
becomes larger, $\eta$ obtains a constant value
for small energies. Thus, in this case of large $S$,
only one asymptotic function is needed.
For smaller values of $S$, there are clearly
two separated branches for $\eta$, and we expect that
two asymptotic functions are required to fully describe
SRCs.
For very small values of $S$, the channels are effectively not coupled,
and we go back to the known non-coupled case. Then, usually, there
is one preferred channel, and again only one function is needed.
The same general picture is obtained for other coupled channels
or if $V_0$ is kept constant.
The dependence of the values of $S$, that can be considered
"large", on the different parameters of the potential
is studied in the supplemental materials \cite{supp}.

%, and look on the two lowest allowed
%positive energies, $E_1$ and $E_2$. We
%compare the corresponding ratios $c_2(E_i)/c_1(E_i)$ for these two energies.
%If the two ratios are similar we can deduce that only one function is needed
%to describe the SRCs in this channel. In Fig. \ref{strngth} we present
%this comparison for different values of $R$. We can see that for large enough
%values of $S$, corresponding to a strong coupling between the two channels,
%only one function is needed. 
%For very small values of $S$, the channels are effectively not coupled,
%and we go back to the known non-coupled case. In this case, usually there
%is one preferred channel, and again only one function will be needed.
%We can see that there is an intermediate range of $S$ for which it seems
%that two functions will be needed to fully describe the SRCs in these
%two coupled channels.
%The same general picture is obtained for other coupled channels
%or if $V_0$ is kept constant.

The conclusion that only one functions is needed in the
strong coupling limit
holds only when considering the zero-energy solutions.
If energy corrections are included, two functions might be required at some point,
since two branches clearly
exist when going to higher energies. 
%
%\begin{figure}\begin{center}
%\includegraphics[width=8.6 cm]{changed_ratio_VS_strength.eps}
%\caption{\label{strngth} 
%Comparison between the values of the ratio
% $\eta(E) \equiv c_2(E)/c_1(E)$ for the first
%two allowed energies, $E_1$ and $E_2$, in a box with radius $R$, using
%the simple potential given by Eqs. \eqref{Gauss1} and \eqref{Gauss2},
%for different values of $S$. We used
%$\Lambda^2=3$ fm$^{-2}$ and $V_0$ was tuned such that the scattering
%length will have a constant value $a=10$ fm.
%We can see that $\eta(E_1) \approx \eta(E_2)$
%if $S$ is large enough.
%The blue, red and green lines
%are for $R=15, 25, 35$ fm.
%\red{(change figure to eta vs. E for a given R,
%and plot it for different values of S that show the trend
%(use normalized wave functions))}
%}
%\end{center}\end{figure}

%The contact theory for coupled channels is also
%relevant for the case of a zero-range interaction.
%Using the S-matrix, and its energy dependence for small energies,
%the 2B wave functions of the asymptotic form can be written explicitly.
%As a result, for the case of coupled $s$-wave and $d$-wave channels, 
%the momentum distribution will include both a $k^{-4}$ term and a constant term
%together (\red{see supplementary materials}).
%The zero-range case can be also studied analytically using a separable
%non-local potential. For such a potential, the asymptotic momentum distribution
%of a shallow two-body bound state can be obtained.
%For the $s$-wave and $d$-wave case, the potential
%should be fine tuned in order to get a significant contribution from the
%$d$-wave constant term (\red{see supplementary materials}).

%=============================================================================
% Summary
%============================================================================= 
{\it Summary -- } 
Summing up, in this work the contact formalism was generalized
to systems with coupled channels.
Focusing on the case of two coupled channels, the relevant asymptotic
form was presented and the matrices of contacts were defined.
The asymptotic form generally includes two different
2B functions.
Looking on the specific case of nuclear systems, it was observed that
only a single 2B function and a single contact are needed to describe the $pn$ SRCs
in the deuteron channel.
This result led to the understanding that a boundary condition
should be imposed on the 2B functions,
possibly representing the effects of the remaining particles in the system.
Indeed, calculating the 2B functions for the nuclear case in a box
gives a single 2B function, which does not depend on the radius of the box.

This is a clear theoretical explanation for the fact that $pn$ SRCs can be described
in leading order using the deuteron wave function.
We have stressed that the known deuteron-channel dominance does not
explain this observation.
We note again that in this paper we do not try to explain the dominance
of the deuteron channel over other possible channels (such as $T=1$
channels). We focus here on the deuteron channel, and explain
why with in this channel a single 2B function is sufficient.

Analyzing a simpler potential, we have shown that the collapse of the asymptotic
wave functions into a single function is not restricted to nuclear physics,
but it is a universal phenomena if the coupling terms of the potential
are very strong or very weak.
This work should be relevant to any system with SRCs which
is dominated by coupled channels.
It can help in understanding the importance of additional coupled channels to nuclear
SRCs, especially since there are no bound states in these channels. It can also be relevant
to ultra-cold atomic systems with short-range interactions.

%==============================================================================
\begin{acknowledgments}
This work was supported by the Pazy Foundation.
\end{acknowledgments}
%==============================================================================

%==============================================================================

\end{document}